\documentclass[english,pra,aps,superscriptaddress,showpacs]{revtex4}
\usepackage[T1]{fontenc}
\usepackage[latin9]{inputenc}
\usepackage{amsmath}
\usepackage{amssymb}
\usepackage{esint}

\makeatletter
\@ifundefined{textcolor}{}
{%
 \definecolor{BLACK}{gray}{0}
 \definecolor{WHITE}{gray}{1}
 \definecolor{RED}{rgb}{1,0,0}
 \definecolor{GREEN}{rgb}{0,1,0}
 \definecolor{BLUE}{rgb}{0,0,1}
 \definecolor{CYAN}{cmyk}{1,0,0,0}
 \definecolor{MAGENTA}{cmyk}{0,1,0,0}
 \definecolor{YELLOW}{cmyk}{0,0,1,0}
 }

\usepackage{mathrsfs}\usepackage{amsfonts}\usepackage{epsfig}\@ifundefined{definecolor}{\@ifundefined{definecolor}
 {\usepackage{color}}{}
}{}\usepackage{CJK}\usepackage{pifont}\usepackage{bbding}

\makeatother

\usepackage{babel}

\makeatother

\usepackage{babel}

\begin{document}

\title{Aharonov-Anandan phases in Lipkin-Meskov-Glick model}

\author{Da-Bao Yang}

\affiliation{Theoretical Physics Division, Chern Institute of Mathematics, Nankai
University, Tianjin 300071, People's Republic of China}

\author{Jing-Ling Chen}

\email{chenjl@nankai.edu.cn}

\affiliation{Theoretical Physics Division, Chern Institute of Mathematics, Nankai
University, Tianjin 300071, People's Republic of China}

\date{\today}
\begin{abstract}
In the system of several interacting spins, geometric phases have
been researched intensively. However, the studies are mainly focused
on the adiabatic case (Berry phase), so it is necessary for us to
study the non-adiabatic counterpart (Aharonov and Anandan phase).
In this paper, we analyze both the non-degenerate and degenerate geometric
phase of Lipkin-Meskov-Glick type model, which has many application
in Bose-Einstein condensates and entanglement theory. Furthermore,
in order to calculate degenerate geometric phases, the Floquet theorem
and decomposition of operator are generalized. And the general formula
is achieved. 
\end{abstract}

\pacs{03.65.Vf, 75.10.Pq, 31.15.ac}

\maketitle

\section{Introduction}

\label{sec:introduction}

Geometric phase relating to quantum mechanics is one of the most interesting
developments in the recent 25 years, which has been discovered by
Berry \cite{berry1984quantal} in the context of adiabatic, unitary,
cyclic evolution of time-dependent quantum system. He demonstrated
that besides the usual dynamical phase, an additional phase relating
to geometry of the state space was generated. Soon Simon \cite{simon1983holonomy}
give an geometrical interpretation of Berry's phase. Berry phase can
be regarded as the holonomy of a line bundle $L$ over the space of
parameters $M$ of the system, if $L$ is endowed with a natural connection.
Subsequently the degenerate case of Berry phase was generalized by
Wilczek and Zee \cite{zee1988nonabelian}.

Discarding the assumption of adiabaticity, Aharonov and Anandan \cite{aharonov1987phase}
generalized Berry's result. The dynamical phase was identified as
the integral of the expectation value of the Hamiltonian. The Aharonov
and Anandan phase (A-A phase) could be obtained by the difference
between the total phase and the dynamical one and also be determined
by the natural connection on a $U(1)$ principle fiber bundle over
the space of projective Hilbert space. Soon after, Anandan \cite{anandan1988nonadiabatic}
generalized the above one to the degenerate case.

Depending on the Pancharatnam's earlier work \cite{shapere1989geometric},
Samuel and Bhandari \cite{samuel1988general} found a more general
phase in the context of non-cyclic and non-unitary evolution of quantum
mechanics. Furthermore, there are some reviewed papers \cite{moore1991calculation,Anandan1992geometric}
and books \cite{shapere1989geometric,bohm2003geometric} about the
theoretical developments, experiments and applications of geometric
phase.

Recently, the study of geometric phase of a composite system of several
spins has attracted a lot of attention. Sj$\ddot{o}$vist \cite{sjoquist2000entangled}
analyzed the non-cyclic and non-adiabatic two-particle geometric phase
for a pair of entangled spins in a time-independent uniform magnetic
field. Tong, Kwek and Oh \cite{tong2003entangled} generalized the
above case and calculated the geometric phase of the similar system
in a rotating magnetic field. Yi, Wang and Zheng \cite{yi2004berry}
investigated the adiabatic and cyclic geometric phase of two coupled
spin-1/2 system, one of which is driven by a varying magnetic field.
Xing \cite{xing2006biprtite,xing2006new} studied further the adiabatic
and cyclic geometric phase of two and three coupled spin-$1/2$ system
with anisotropic interactions. Shi \cite{shi2008geometric,shi2010unitary}
researched the Berry and A-A phases of two Heisenberg-coupled and
Ising-coupled qubits in a rotating field respectively, both of which
had a significance in quantum computing application and measuring
geometric phases. Lately Sj$\ddot{o}$vist, $\emph{\emph{et. al.}}$
\cite{sjoqvist2010berry}, analyzed the adiabatic geometric phase
of ground state of finite-size Lipkin-Meskov-Glick type model (LMG)
which consists of three spin-$\frac{1}{2}$ particles. In this paper,
we calculate the non-adiabatic and cyclic geometric phase, namely
Aharonov and Anandan phase, of this system.

The outline of the present paper is as follows. Section II. reviews
the non-degenerate and degenerate Aharonov and Anandan phase. And
a method of calculation for A-A phase is introduced in order to calculate
the LMG model. Moreover, we generalize the above methods to the degenerate
case. In Sec. III., the cyclic state of the LMG model is solved. Furthermore,
the Aharonov and Anandan phase for non-degenerate and degenerate case
are calculated respectively. In Sec. IV., the non-degenerate adiabatic
phases, namely Berry phases, are analyzed. And the connection between
A-A phase and Berry phase is discussed according to quantum adiabatic
theorem. At last, a conclusion was drawn.

\section{Reviews Of A-A Phase And its generalization to degeneracy case}

\label{sec:reviews}

Consider a general quantum system with time-dependent Hamiltonian
$H(t)$, which is $T$- period, i.e., $H(T)=H(0)$. Given an initial
state of the system $\psi(0)$, The evolution is determined by Schr$\ddot{o}$inger
equation, that is, \begin{equation}
i\frac{d}{dt}|\psi(t)\rangle=H(t)|\psi(t)\rangle.\label{eq:Dyanmical}\end{equation}
 Choosing an initial state makes the evolution cyclic, i.e., $\psi(T)=e^{i\chi}\psi(0)$,
where $\chi$ is the overall phase. Furthermore, the overall phase
can be split into two parts, namely the dynamical phase and the geometric
phase \cite{aharonov1987phase,moore1991calculation,bohm2003geometric}.
The dynamical phase has a natural definition, which is\begin{equation}
\delta=-\int_{0}^{T}\langle\psi(t)|H(t)|\psi(t)\rangle dt.\label{eq:DyanicalPhase}\end{equation}
 Hence \begin{equation}
\eta=\chi-\delta\label{eq:GeometricPhase}\end{equation}
 is the geometric phase, which is determined by purely geometric property
of evolution. In order to uncover the mystery of the geometric phase,
we introduce a single-valued vector \begin{equation}
|\tilde{\psi}(t)\rangle=e^{-i\theta(t)}|\psi(t)\rangle\label{eq:SingleValuedVector}\end{equation}
 such that $\theta(0)=0$ and $\theta(T)=\chi$. Substituting Eq.
\eqref{eq:Dyanmical} and Eq. \eqref{eq:SingleValuedVector} into
Eq. \eqref{eq:DyanicalPhase}, the dynamical can be rephrased into\[
\delta=\chi-i\int_{0}^{T}\langle\tilde{\psi}|\frac{d}{dt}|\tilde{\psi}\rangle dt.\]
 By substitution the above Equation into Eq. \eqref{eq:GeometricPhase},
one gets\begin{equation}
\eta=i\int_{0}^{T}\mathcal{A}dt,\label{eq:NondegenerateAAPhase}\end{equation}
 where $\mathcal{A}=\langle\tilde{\psi}|\frac{d}{dt}|\tilde{\psi}\rangle$
The above formula is gauge invariant and also has a mathematical counterpart
called holonomy in $U(1)$ principle fiber bundle whose base manifold
is the projective Hilbert space.

In the previous paragraph, the non-degenerate Aharonov and Anandan
phase is elucidated, moreover, let's talk about the its generalization,
which is call the degenerate A-A phase \cite{anandan1988nonadiabatic,bohm2003geometric}.
As is known, the state vectors live in Hilbert space $H$. Now, we
focus on the subspace $V_{n}(t)$ of $H$, where $n(n>1)$ is the
notation of dimension of the subspace. Furthermore, $V_{n}(t)$ undergoes
cyclic evolution, such that $V_{n}(T)=V_{n}(0)$. Let $\{|\tilde{\psi}_{n\alpha}(t)\rangle,\alpha=1,\cdots,f_{n}\}$
be an orthonormal basis of $V_{n}(t)$ with $|\tilde{\psi}_{\alpha}(T)\rangle=|\tilde{\psi}_{\alpha}(0)\rangle$
for every $\alpha$, where we write $|\tilde{\psi}_{\alpha}(t)\rangle$
instead of $|\tilde{\psi}_{n\alpha}(t)\rangle$ for short. By comparison
with the non-degenerate case, it is easy to deduced that $|\tilde{\psi}_{\alpha}(t)\rangle$
is degenerate generalization of of the single valued vector $|\tilde{\psi}(t)\rangle$.
With the initial state $|\psi_{\alpha}(0)\rangle=|\tilde{\psi}_{\alpha}(0)\rangle$,
substituting\[
|\psi_{\alpha}(t)\rangle=\sum_{\beta=1}^{n}U_{\beta\alpha}(t)|\tilde{\psi}_{\beta}(t)\rangle\]
 into the Schr$\ddot{o}$inger Eq. \eqref{eq:Dyanmical}, we get\begin{equation}
U(t)=\mathcal{T}exp\{\int_{0}^{T}i[\mathcal{A}(t)-\mathcal{E}(t)]dt\},\label{eq:TimeEvolutionOperator}\end{equation}
 where $\mathcal{A}_{\alpha\beta}=\langle\tilde{\psi}_{\alpha}(t)|\frac{d}{dt}|\tilde{\psi}_{\beta}(t)\rangle$
and $\mathcal{E}_{\alpha\beta}=\langle\tilde{\psi}_{\alpha}(t)|H(t)|\tilde{\psi}_{\beta}(t)\rangle.$
As the matrix $\mathcal{A}$ and $\mathcal{E}$ do not generally commute,
the total $U(T)$ can't be written as the product of the dynamical
phase,\begin{equation}
U^{dynamical}=\mathcal{T}exp(-i\int_{0}^{T}\mathcal{E}(t)dt),\label{eq:DegenerateDynamicalPhase}\end{equation}
 and the geometrical phase,\begin{equation}
U^{geometric}=\mathcal{T}exp(i\int_{0}^{T}\mathcal{A}(t)dt).\label{eq:DegenerateAAPhaseTime}\end{equation}
 However, the latter quantity can be transformed into the path-order
integral,\begin{equation}
U^{geometric}=\mathcal{P}exp(i\oint_{\mathcal{C}}\mathcal{A}(t)dt).\label{eq:DegenerateAAPhasePath}\end{equation}
 Moreover, it can be regarded as the holonomy of $U(N)$ principle
fiber bundle with a natural connection, whose base space is the grassmann
manifold.

From the above elucidation, to calculate the Aharonov and Anandan
phase is to find the cyclic initial state and the single-valued vector.
Now we concentrate on a periodic Hamiltonian $H(t)$ with period $T$.
From the well-known Floquet theory, the time evolution operation $U(t)$
can be written as\begin{equation}
U(t)=Z(t)e^{iMt},\label{eq:DecompositionFromFloquetTheorom}\end{equation}
 where $Z(t)$ is a unitary $T$-period operation with $Z(0)=1$ and
$M$ is a Hermitian operator. Moore and Stedman \cite{stedman1990nonadiabatic,moore1991calculation}
had discussed application of the above result into the non-degenerate
case. Specifically speaking, the connection one-form can be expressed
as\begin{equation}
\mathcal{A}=i\langle n|Z^{\dagger}(t)\frac{d}{dt}Z(t)|n\rangle,\label{eq:ConcretNondegenerateConnection}\end{equation}
 where $|n\rangle$ is the $n$th eigenvector of $M$, which is also
the cyclic initial state, hence $Z(t)|n\rangle$ is the single valued
vector. Moreover, we point out that this theorem can also be used
to calculate the degenerate case. For simplicity, suppose that $M$
has $m$th eigenvalue, which is degenerate and spanned up a $f_{m}$
subspace with orthonormal basis, and the the eigenvectors are expressed
as $|m\alpha\rangle$. It is very easy to verify that the $\{|m\alpha\rangle,\alpha=1,2,\cdots,f_{m}\}$
are cyclic initial states and single-valued vector are $Z(t)|m\alpha\rangle,\alpha=1,2,\cdots,f_{m}\}$.
Hence, we can calculate the connection matrix of degenerate case,
which is\begin{equation}
\mathcal{A}_{\alpha\beta}=i\langle m\alpha|Z^{\dagger}(t)\frac{d}{dt}Z(t)|m\beta\rangle.\label{eq:ConcretDegenerateConnection}\end{equation}
 Thus, substituting Eq. \eqref{eq:ConcretNondegenerateConnection}
and Eq. \eqref{eq:ConcretDegenerateConnection} into Eq. \eqref{eq:NondegenerateAAPhase}
and Eq. \eqref{eq:DegenerateAAPhaseTime} respectively, both the non-degenerate
and degenerate A-A phase can be obtained.

However, the time evolution operator is very hard to get according
to Eq. \eqref{eq:TimeEvolutionOperator}.So generally we can't follow
the procedure which is displayed in the previous paragraph to calculate
the A-A phase.Nevertheless, don't be so discouraged and let's consider
an important time-dependent Hamiltonian which has this form\begin{equation}
H(t)=e^{-iAt}\tilde{H}e^{iAt},\label{eq:Hamiltonian}\end{equation}
 where $A$ and $\tilde{H}$ are time independent. Substituting Eq.
\eqref{eq:Hamiltonian} into Eq. \eqref{eq:Dyanmical},then multiplying
$e^{iAt}$ at both side of the equation, one can get\[
ie^{iAt}\frac{d|\psi\rangle}{dt}=\tilde{H}e^{iAt}|\psi\rangle.\]
 By use of the derivative formula $e^{iAt}d|\psi\rangle/dt=d(e^{iAt}|\psi\rangle)/dt-|\psi\rangle de^{iAt}/dt$,
it is not very difficult to see that the time evolution operation
can be written as\begin{equation}
U(t)=e^{-iAt}e^{-iBt},\label{eq:DecompositionFromTPeriodHamiltonian}\end{equation}
 where $B=\tilde{H}-A$ \cite{salzman1974quantum,moore1992berry,moore1991calculation}.
Again, suppose that the Hamiltonian is $H$ is $T$-period. And it
is a sufficient and necessary condition that $e^{-iAT}$ commutes
with B. Hence, they have a complete set of simultaneous eigenvectors,
i.e.,\begin{equation}
B\phi_{n}=B_{n}\phi_{n},\label{eq:BEigenEquation}\end{equation}
 \begin{equation}
e^{-iAT}\phi_{n}=e^{-i\theta_{n}}\phi_{n},\label{eq:AEigenEquation}\end{equation}
 where we suppose that $B$ have non-degenerate eigenvalues. This
case had already discussed by Moore \cite{moore1992berry,moore1991calculation}.
Its A-A phase \cite{moore1992berry,moore1991calculation} is\begin{equation}
\eta_{n}=\langle\phi_{n}|A|\phi_{n}\rangle T-\theta_{n}.\label{eq:AAPhaseOfNondegenerateLMG}\end{equation}

From above, we can see that the calculational methods for the non-degenerate
A-A phase was displayed. Moreover, there exists many degenerate A-A
phase in actual physical system. Nevertheless, few physicists consider
about this problem, except Mostafazadeh \cite{mostafazadeh1998nonabelian}.
He uses the methods of dynamical invariants to obtain the degenerate
geometric phase. But now let us fallow another line which is more
direct and convenient and generalize it to the degenerate case. we
still confer to the non-degenerate case which is depicted above, so
Eq. \eqref{eq:BEigenEquation} and Eq. \eqref{eq:AEigenEquation}
can be expressed as\[
B\phi_{n\alpha}=b_{n}\phi_{n\alpha},\]
 \[
e^{-iAT}\phi_{n\alpha}=e^{-i\theta_{n}}\phi_{n\alpha},\]
 where $\alpha=1,2,\cdots f_{n}$ and the degenerate space is a $f_{n}$-fold
subspace. Subsequently, I want to transform Eq. \eqref{eq:DecompositionFromTPeriodHamiltonian}
into the formula of Eq. \eqref{eq:DecompositionFromFloquetTheorom}.
However, $e^{-iAt}$ isn't generally $T$-periodic, so we introduce
an operator $\Omega$ which is commute with $B$. Hence, they have
the simultaneous eigenvectors, so we can get\begin{equation}
\Omega\phi_{n\alpha}=\omega_{n}\phi_{n\alpha}.\label{eq:OmegaEigenEquation}\end{equation}
 Furthermore, let $\Omega$ still satisfy the following properties:\[
Z(t)=e^{-iAt}e^{i\Omega t/T},\]
 \[
M=-B-\Omega/T.\]
 So the single-valued vector becomes $e^{-iAt}e^{i\Omega t/T}|\phi_{n\alpha}\rangle$.
Thus we can get\begin{equation}
\mathcal{A}_{\alpha\beta}=\langle\phi_{n\alpha}|A|\phi_{n\beta}\rangle-\frac{\omega_{n}}{T}\delta_{\alpha\beta}.\label{eq:ConnectionOfDegenerateCase}\end{equation}
 Substituting the above Eq. \eqref{eq:ConnectionOfDegenerateCase}
into Eq. \eqref{eq:DegenerateAAPhaseTime} or Eq. \eqref{eq:DegenerateAAPhasePath},
we can get the corresponding A-A phase. Another quantity can also
be achieved by the similar calculation, i.e.,\begin{equation}
\mathcal{E}_{\alpha\beta}=\langle\phi_{n\alpha}|\tilde{H}|\phi_{n\beta}\rangle.\label{eq:DynamicalOneForm}\end{equation}
 Hence, the dynamical phase can be obtained by substitution Eq. \eqref{eq:DynamicalOneForm}
into Eq. \eqref{eq:DegenerateDynamicalPhase}.

\section{The Aharonov and Anandan Phase Of LMG}

\label{sec:Nonadiabatic}

The previous section has introduced methods to calculate the non-degenerate
and degenerate A-A phase. The methods will be use to calculate a concrete
model called three qubits LMG \cite{sjoqvist2010berry} in this section.
Its Hamiltonian reads \[
\tilde{H}=-\frac{1}{3}(S_{x}^{2}+\gamma S_{y}^{2})-hS_{z},\]
 where $\gamma$ is an anisotropy parameter, $h$ is the strength
of an external magnetic field along the $z$ direction, and $S_{\alpha}=\frac{1}{2}\sum_{k=1}^{N}\sigma_{\alpha}^{k}$
is the $\alpha$th component of the spin operator(for simplicity,
we set $\hbar=1$ from now on) with $\sigma_{x}^{k}$, $\sigma_{y}^{k}$
and $\sigma_{z}^{k}$ are Pauli operators of the $k$the qubit in
the representation of $\sigma_{z}$ \cite{sjoqvist2010berry}. Ignoring
the trivial and constant term $-\frac{1}{4}(1+\gamma)$ of $\tilde{H}$,
the Hamiltonian becomes\[
\tilde{H}=-\frac{1}{6}[\sigma_{x}^{1}\sigma_{x}^{2}+\sigma_{x}^{2}\sigma_{x}^{3}+\sigma_{x}^{1}\sigma_{x}^{3}+\gamma(\sigma_{y}^{1}\sigma_{y}^{2}+\sigma_{y}^{2}\sigma_{y}^{3}+\sigma_{y}^{1}\sigma_{y}^{3})]-\frac{h}{2}(\sigma_{z}^{1}+\sigma_{z}^{2}+\sigma_{z}^{3}).\]
 Moreover, let's consider about the isospectral one-parameter Hamiltonian
family\begin{equation}
H=e^{-i\phi S_{z}}\tilde{H}e^{i\phi S_{z}},\label{eq:HamiltonianOfLMG}\end{equation}
whose explicit expression takes this form\[
\begin{array}{c}
-\frac{1}{6}[(\cos^{2}\phi+\gamma\sin^{2}\phi)(\sigma_{x}^{1}\sigma_{x}^{2}+\sigma_{x}^{2}\sigma_{x}^{3}+\sigma_{x}^{1}\sigma_{x}^{3})+(\sin^{2}\phi+\gamma\cos^{2}\phi)(\sigma_{y}^{1}\sigma_{y}^{2}+\sigma_{y}^{2}\sigma_{y}^{3}+\sigma_{y}^{1}\sigma_{y}^{3})\\
+(1-\gamma)\sin\phi\cos\phi(\sigma_{x}\sigma_{y}-\sigma_{x}^{1}\sigma_{y}^{1}-\sigma_{x}^{2}\sigma_{y}^{2}-\sigma_{x}^{3}\sigma_{y}^{3})]-\frac{h}{2}(\sigma_{z}^{1}+\sigma_{z}^{2}+\sigma_{z}^{3})\end{array},\]
where $\phi=\omega t$ is a varying parameter and $\omega$ is the
angular velocity. By a glance at Eq. \eqref{eq:Hamiltonian}, a conclusion
can be drawn that the two Hamiltonian have a similar form. Hence,
we can take the similar procedure to calculate the corresponding A-A
phase. The time evolution operator becomes\begin{equation}
U(t)=e^{-iAt}e^{-iBt},\label{eq:TimeEvolutionOperatorOfLMG}\end{equation}
 where $A=\omega S_{z}$ and $B=\tilde{H}-\omega S_{z}$. Next, we
intend to calculate the cyclic initial state which is the eigenvectors
of $B$ in this case. Thus, at first, the operator $B$ must be represented
in a concrete basis, which are $\{|000\rangle,|011\rangle,|101\rangle,|110\rangle,|111\rangle,|100\rangle,|010\rangle,|001\rangle\}$
, where $|0\rangle$ represents spin up and $|1\rangle$ represents
spin down. So $B$ takes the block-diagonal form\[
B(\gamma,h,\omega)=\left(\begin{array}{cc}
P(\gamma,h,\omega) & 0\\
0 & P(\gamma,-h,-\omega)\end{array}\right),\]
 where\[
P(\gamma,h,\omega)=\left(\begin{array}{cccc}
-\frac{3}{2}(h+\omega) & -\frac{1}{6}(1-\gamma) & -\frac{1}{6}(1-\gamma) & -\frac{1}{6}(1-\gamma)\\
-\frac{1}{6}(1-\gamma) & \frac{1}{2}(h+\omega) & -\frac{1}{6}(1+\gamma) & -\frac{1}{6}(1+\gamma)\\
-\frac{1}{6}(1-\gamma) & -\frac{1}{6}(1+\gamma) & \frac{1}{2}(h+\omega) & -\frac{1}{6}(1+\gamma)\\
-\frac{1}{6}(1-\gamma) & -\frac{1}{6}(1+\gamma) & -\frac{1}{6}(1+\gamma) & \frac{1}{2}(h+\omega)\end{array}\right),\]
 and $0$ is the $4\times4$ null matrix. Because $B(\gamma,h,\omega)=P(\gamma,h,\omega)\oplus P(\gamma,-h,-\omega)$,
we can investigate $P(\gamma,h,\omega)$ and $P(\gamma,-h,-\omega)$
respectively. However, the similar information from $P(\gamma,-h,-\omega)$
can be obtain from $P(\gamma,h,\omega)$. So we can focus on the subspace
of solutions which the sub-matrix $P(\gamma,h,\omega)$ acts to simplify
the problem. Thus the two eigenvalues of $P(\gamma,h,\omega)$ are\[
p_{1}=-\frac{1}{2}(\omega+h)-\frac{1}{6}(1+\gamma)-\frac{1}{3}\sqrt{r},\]
 \[
p_{2}=-\frac{1}{2}(\omega+h)-\frac{1}{6}(1+\gamma)+\frac{1}{3}\sqrt{r}\]
 and\[
p_{3}=p_{4}=\frac{1}{2}(h+\omega)+\frac{1}{6}(1+\gamma),\]
 where\[
r=9h^{2}+9\omega^{2}+\gamma^{2}+18h\omega-3h\gamma-3\gamma\omega-3\omega-3h-\gamma+1.\]
 From above calculation, we can draw a conclusion that $p_{1}$ and
$p_{2}$ correspond to the non-degenerate case whereas $p_{3}=p_{4}$
corresponds to the degenerate case. However, its A-A phase factor
is proved to be trivial. So we only focus on the non-degenerate case.
And the corresponding two orthogonal eigenvectors are \begin{equation}
|\phi_{1}\rangle=\frac{1}{\sqrt{n_{1}}}\left(\begin{array}{cccc}
\gamma+1-6(\omega+h)-2\sqrt{r} & \gamma-1 & \gamma-1 & \gamma-1\end{array}\right)^{T}\label{eq:Eigenvector1}\end{equation}
 and \begin{equation}
|\phi_{2}\rangle=\frac{1}{\sqrt{n_{2}}}\left(\begin{array}{cccc}
\gamma+1-6(\omega+h)+2\sqrt{r} & \gamma-1 & \gamma-1 & \gamma-1\end{array}\right)^{T},\label{eq:Eigenvector2}\end{equation}
 where\[
n_{1}=3(\gamma-1)^{2}+[-6(\omega+h)+\gamma+1-2\sqrt{r}]^{2},\]
 \[
n_{2}=3(\gamma-1)^{2}+[-6(\omega+h)+\gamma+1+2\sqrt{r}]^{2},\]
 and $T$ denotes the transpose operation of matrix.

We have already calculated the cyclic initial state of the system.
Moreover, from Eq. \eqref{eq:HamiltonianOfLMG}, it is very easy to
verify that the system has $2\pi/\omega$-periodic Hamiltonian. In
the following paragraphs, we will calculated the corresponding A-A
phase of the non-degenerate case and degenerate case.

Substituting $T=2\pi/\omega$ into $e^{-i\omega TS_{z}}$, by use
of Eq. \eqref{eq:AEigenEquation}, we can choose that \begin{equation}
\theta_{n}=\pi.\label{eq:ThetaNondegenerate}\end{equation}
 Subsequently, we represent $A$ in the given basis, which is expressed
as\begin{equation}
A=\frac{1}{2}\omega\left(\begin{array}{cccc}
3 & 0 & 0 & 0\\
0 & -1 & 0 & 0\\
0 & 0 & -1 & 0\\
0 & 0 & 0 & -1\end{array}\right).\label{eq:AOfLMG}\end{equation}
 Substituting the above Eq. \eqref{eq:ThetaNondegenerate}, Eq. \eqref{eq:AOfLMG}
and Eq. \eqref{eq:Eigenvector1} into Eq. \eqref{eq:AAPhaseOfNondegenerateLMG},
one can get the A-A phase corresponding to the first cyclic initial
state, which reads\begin{equation}
\eta_{1}=\frac{3\pi}{n_{1}}\{[1+\gamma-6(\omega+h)-2\sqrt{r}]^{2}-(1-\gamma)^{2}\}-\pi.\label{eq:A-APhase1}\end{equation}
 By the similar procedure, the second geometric phase is\begin{equation}
\eta_{2}=\frac{3\pi}{n_{2}}\{[1+\gamma-6(\omega+h)+2\sqrt{r}]^{2}-(1-\gamma)^{2}\}-\pi.\label{eq:A-APhase2}\end{equation}

The above calculation is all about A-A phase of the non-degenerate
case. Next, A-A phase of the degenerate case will be discussed. Nevertheless,
the degenerate one is trivial one, so it is necessary to reconsider
another system whose Hamiltonian is \[
H=e^{-i\phi S_{x}}\tilde{H}e^{i\phi S_{x}}.\]
 According to the operator decomposition, the time evolution operator
is\[
U(t)=e^{-iAt}e^{-iBt},\]
 where $A=\omega S_{x}$ and $B=\tilde{H}-\omega S_{x}.$ Moreover,
$B$ can be represented by the given basis as\[
B=\left(\begin{array}{cccccccc}
-\frac{3}{2}h & \frac{1}{6}(\gamma-1) & \frac{1}{6}(\gamma-1) & \frac{1}{6}(\gamma-1) & 0 & -\frac{\omega}{2} & -\frac{\omega}{2} & -\frac{\omega}{2}\\
\frac{1}{6}(\gamma-1) & \text{\ensuremath{\frac{h}{2}}} & -\frac{1}{6}(\gamma+1) & -\frac{1}{6}(\gamma+1) & -\frac{\omega}{2} & 0 & -\frac{\omega}{2} & -\frac{\omega}{2}\\
\frac{1}{6}(\gamma-1) & -\frac{1}{6}(\gamma+1) & \text{\ensuremath{\frac{h}{2}}} & -\frac{1}{6}(\gamma+1) & -\frac{\omega}{2} & -\frac{\omega}{2} & 0 & -\frac{\omega}{2}\\
\frac{1}{6}(\gamma-1) & -\frac{1}{6}(\gamma+1) & -\frac{1}{6}(\gamma+1) & \text{\ensuremath{\frac{h}{2}}} & -\frac{\omega}{2} & -\frac{\omega}{2} & -\frac{\omega}{2} & 0\\
0 & -\frac{\omega}{2} & -\frac{\omega}{2} & -\frac{\omega}{2} & \frac{3}{2}h & \frac{1}{6}(\gamma-1) & \frac{1}{6}(\gamma-1) & \frac{1}{6}(\gamma-1)\\
-\frac{\omega}{2} & 0 & -\frac{\omega}{2} & -\frac{\omega}{2} & \frac{1}{6}(\gamma-1) & \text{-\ensuremath{\frac{h}{2}}} & -\frac{1}{6}(\gamma+1) & -\frac{1}{6}(\gamma+1)\\
-\frac{\omega}{2} & -\frac{\omega}{2} & 0 & -\frac{\omega}{2} & \frac{1}{6}(\gamma-1) & -\frac{1}{6}(\gamma+1) & -\frac{h}{2} & -\frac{1}{6}(\gamma+1)\\
-\frac{\omega}{2} & -\frac{\omega}{2} & -\frac{\omega}{2} & 0 & \frac{1}{6}(\gamma-1) & -\frac{1}{6}(\gamma+1) & -\frac{1}{6}(\gamma+1) & -\frac{h}{2}\end{array}\right).\]
 Now we only focus on its two degenerate eigenvalues which are \[
B_{1}=\frac{1}{6}(1+\gamma-3\sqrt{h^{2}+\omega^{2}}),\]
 and\[
B_{2}=\frac{1}{6}(1+\gamma+3\sqrt{h^{2}+\omega^{2}}).\]
 And the corresponding eigenvectors are\begin{equation}
|\phi_{11}\rangle=\frac{1}{\sqrt{N_{11}}}\left(\begin{array}{cccccccc}
0 & \frac{1}{\omega}(\sqrt{h^{2}+\omega^{2}}-h) & 0 & \frac{1}{\omega}(h-\sqrt{h^{2}+\omega^{2}}) & 0 & -1 & 0 & 1\end{array}\right)^{T},\label{eq:11InitialCyclicStateOfSx}\end{equation}

\begin{equation}
|\phi_{12}\rangle=\frac{1}{\sqrt{N_{12}}}\left(\begin{array}{cccccccc}
0 & \frac{1}{2\omega}(\sqrt{h^{2}+\omega^{2}}-h) & \frac{1}{\omega}(h-\sqrt{h^{2}+\omega^{2}}) & \frac{1}{2\omega}(\sqrt{h^{2}+\omega^{2}}-h) & 0 & -\frac{1}{2} & 1 & -\frac{1}{2}\end{array}\right)^{T},\label{eq:12InitialCyclicStateOfSx}\end{equation}
 and\[
|\phi_{21}\rangle=\frac{1}{\sqrt{N_{21}}}\left(\begin{array}{cccccccc}
0 & -\frac{1}{\omega}(h+\sqrt{h^{2}+\omega^{2}}) & 0 & \frac{1}{\omega}(h+\sqrt{h^{2}+\omega^{2}}) & 0 & -1 & 0 & 1\end{array}\right)^{T},\]
 \[
|\phi_{22}\rangle=\frac{1}{\sqrt{N_{22}}}\left(\begin{array}{cccccccc}
0 & -\frac{1}{2\omega}(h+\sqrt{h^{2}+\omega^{2}}) & \frac{1}{\omega}(h+\sqrt{h^{2}+\omega^{2}}) & -\frac{1}{2\omega}(h+\sqrt{h^{2}+\omega^{2}}) & 0 & -\frac{1}{2} & 1 & -\frac{1}{2}\end{array}\right)^{T},\]
 where $N_{11}=\frac{4}{1+h/\sqrt{h^{2}+\omega^{2}}}$, $N_{12}=\frac{3}{1+h/\sqrt{h^{2}+\omega^{2}}}$,
$N_{21}=4+\frac{4h(h+\sqrt{h^{2}+\omega^{2}})}{\omega^{2}}$, and
$N_{22}=3+\frac{3h(h+\sqrt{h^{2}+\omega^{2}})}{\omega^{2}}$.

It is very easy to verify that $e^{-i\frac{1}{2}\sigma_{x}\omega t}$
is not a $2\pi/\omega$ period unitary operator, but $e^{-i\frac{1}{2}\sigma_{x}\omega t}e^{i\Omega t/T}$
is so, where \begin{equation}
\Omega=\pi\left(\begin{array}{cc}
1 & 0\\
0 & 1\end{array}\right).\label{eq:Omega2}\end{equation}
 In order to calculate the degenerate connection matrix according
to Eq. \eqref{eq:ConnectionOfDegenerateCase}, we must represent $A$
in a matrix form, that is\begin{equation}
A=\frac{1}{2}\omega\left(\begin{array}{cccccccc}
0 & 0 & 0 & 0 & 0 & 1 & 1 & 1\\
0 & 0 & 0 & 0 & 1 & 0 & 1 & 1\\
0 & 0 & 0 & 0 & 1 & 1 & 0 & 1\\
0 & 0 & 0 & 0 & 1 & 1 & 1 & 0\\
0 & 1 & 1 & 1 & 0 & 0 & 0 & 0\\
1 & 0 & 1 & 1 & 0 & 0 & 0 & 0\\
1 & 1 & 0 & 1 & 0 & 0 & 0 & 0\\
1 & 1 & 1 & 0 & 0 & 0 & 0 & 0\end{array}\right).\label{eq:AOfSx}\end{equation}
 Substituting Eq. \eqref{eq:11InitialCyclicStateOfSx}, Eq. \eqref{eq:12InitialCyclicStateOfSx},
Eq. \eqref{eq:AOfSx} and Eq. \eqref{eq:Omega2} into Eq. \eqref{eq:ConnectionOfDegenerateCase},
the connection matrix becomes\begin{equation}
\mathcal{A}=\frac{\omega}{2}(\frac{\omega}{\sqrt{h^{2}+\omega^{2}}}-1)\left(\begin{array}{cc}
1 & 0\\
0 & 1\end{array}\right).\label{eq:ConnectionOfX}\end{equation}
 Further, by substitution Eq. \eqref{eq:ConnectionOfX} into Eq. \eqref{eq:DegenerateAAPhaseTime},
one can get the degenerate A-A phase factor\[
U^{Geometric}=\exp[i\pi(\frac{\omega}{\sqrt{h^{2}+\omega^{2}}}-1)]\left(\begin{array}{cc}
1 & 0\\
0 & 1\end{array}\right).\]
 By the similar procedure, the second degenerate A-A phase factor
reads\[
U^{Geometric}=\exp[-i\pi(\frac{\omega}{\sqrt{h^{2}+\omega^{2}}}+1)]\left(\begin{array}{cc}
1 & 0\\
0 & 1\end{array}\right).\]
 The A-A phase factor have obtained already. Furthermore, we can still
get the dynamical phase factor according to Eq. \eqref{eq:DynamicalOneForm}
and Eq. \eqref{eq:DegenerateDynamicalPhase}.

\section{Discussion And Conclusion }

\label{sec:discussion}

In previous section, we have calculated the non-degenerate and degenerate
A-A phase respectively. For integrity, it is necessary for us to concentrate
on the Adiabatic phase, namely Berry phase, though Sj$\ddot{o}$vist
et. al. \cite{sjoqvist2010berry} had already calculated the Berry
phase of the first model. But their paper probably has a minor bug
of the choice of the instantaneous eigenvectors, which I will explain
in the following paragraphs. Moreover their article only focus on
the energetic ground state of LMG. So it is necessary for us to calculated
the Berry phase for every eigenstates. But the degenerate Berry phase
factor is still trivial, so we only concentrate on the other non-degenerate
Berry phase. The Hamiltonian \eqref{eq:HamiltonianOfLMG} we focus
on have a unified form expressed by Eq. \eqref{eq:Hamiltonian}. According
Berry \cite{berry1984quantal}, Wilczek and Zee \cite{zee1988nonabelian},
in order to determine the Berry phase, we must calculate the instantaneous
eigenstate first. The eigen equation reads\begin{equation}
e^{-iAt}\tilde{H}e^{iAt}|n\rangle=\lambda_{n}|n\rangle.\label{eq:InstantaneousEigenEquation}\end{equation}
 It is difficult to solve the above Equation, the equation can be
transformed\begin{equation}
\tilde{H}|n\rangle^{\prime}=\lambda_{n}|n\rangle^{\prime},\label{eq:TransformedEigenEquation}\end{equation}
 where $|n\rangle^{\prime}=e^{iAt}|n\rangle$. For the LMG model,
the above expression can be transformed to be $|n\rangle=e^{-i\frac{1}{2}\sigma_{z}\omega t}|n\rangle^{\prime}$.
In order to calculate Berry phase, we must choose the instantaneous
eigenvectors to satisfy $|n(T)\rangle=|n(0)\rangle$. However, the
condition can't be satisfied when the period $T=2\pi/\omega$. After
a minor modification, the eigenvector can be chosen to be \begin{equation}
|n\rangle=e^{-i\frac{1}{2}\sigma_{z}\omega}e^{i\pi t/T}|n\rangle^{\prime},\label{eq:eigenvector}\end{equation}
 which is still satisfy Eq. \eqref{eq:InstantaneousEigenEquation}.
Next, we can calculate the non-degenerate connection one-form according
to Berry \cite{berry1984quantal}, which is expressed as\begin{equation}
\mathcal{A}=i\langle n|\frac{d}{dt}|n\rangle.\label{eq:ConnectionOneFormOfBerry}\end{equation}
 Substituting Eq. \eqref{eq:eigenvector} into Eq. \eqref{eq:ConnectionOneFormOfBerry},
one can get\[
\mathcal{A}=\frac{1}{2}\omega\langle n|^{\prime}\sigma_{z}|n\rangle^{\prime}-\frac{\pi}{T}.\]
 Thus the Berry phase becomes\begin{equation}
\eta_{n}=\int_{0}^{T}\mathcal{A}dt.\label{eq:BerryPhaseOfNondegenerate}\end{equation}
 According to Wilczek and Zee \cite{zee1988nonabelian}, by similar
calculations, the degenerate Berry phase factor can be expressed as\begin{equation}
U^{Geometric}=\mathcal{T}exp(i\int_{0}^{T}\mathcal{A}(t)dt),\label{eq:BerryPhaseOfDegenerate}\end{equation}
 where $\mathcal{A}_{\alpha\beta}=i\langle n\alpha|\frac{d}{dt}|n\beta\rangle=\frac{1}{2}\omega\langle n\alpha|^{\prime}\sigma_{z}|m\beta\rangle^{\prime}-\frac{\pi}{T}\delta_{\alpha\beta}$.
At first let us represent $\tilde{H}$ in the given basis, which is\[
\tilde{H}=\left(\begin{array}{cccc}
-\frac{3}{2}h & \frac{1}{6}(\gamma-1) & \frac{1}{6}(\gamma-1) & \frac{1}{6}(\gamma-1)\\
\frac{1}{6}(\gamma-1) & \frac{1}{2}h & -\frac{1}{6}(\gamma+1) & -\frac{1}{6}(\gamma+1)\\
\frac{1}{6}(\gamma-1) & -\frac{1}{6}(\gamma+1) & \frac{1}{2}h & -\frac{1}{6}(\gamma+1)\\
\frac{1}{6}(\gamma-1) & -\frac{1}{6}(\gamma+1) & -\frac{1}{6}(\gamma+1) & \frac{1}{2}h\end{array}\right).\]
 It is not very complicated to get the eigenvalues\[
\lambda_{1}=-\frac{1}{6}(1+\gamma+3h+2\sqrt{q}),\]
 \[
\lambda_{2}=-\frac{1}{6}(1+\gamma+3h-2\sqrt{q}),\]
 and the corresponding eigenvectors are\begin{equation}
|1\rangle=\frac{1}{\sqrt{N_{1}}}\left(\begin{array}{cccc}
1+\gamma-6h-2\sqrt{q} & \gamma-1 & \gamma-1 & \gamma-1\end{array}\right)^{T},\label{eq:v1}\end{equation}
 \begin{equation}
|2\rangle=\frac{1}{\sqrt{N_{2}}}\left(\begin{array}{cccc}
1+\gamma-6h+2\sqrt{q} & \gamma-1 & \gamma-1 & \gamma-1\end{array}\right)^{T},\label{eq:v2}\end{equation}
 where $q=9h^{2}+\gamma^{2}-3h\gamma-3h-\gamma+1$, $N_{1}=3(\gamma-1)^{2}+(1+\gamma-6h-2\sqrt{q})$
and $N_{2}=3(\gamma-1)^{2}+(1+\gamma-6h+2\sqrt{q})$. By the way,
we ignore the degenerate eigenvalues for simplicity as well as for
the triviality of the degenerate case of this system. Substituting
Eq. \eqref{eq:v1} into Eq. \eqref{eq:BerryPhaseOfNondegenerate},
we can get\begin{equation}
\eta_{1}=\frac{3\pi}{N_{1}}[(1+\gamma-6h-2\sqrt{q})^{2}-(\gamma-1)^{2}]-\pi.\label{eq:BerryPhase1}\end{equation}
 Similarly, we can also get\begin{equation}
\eta_{2}=\frac{3\pi}{N_{2}}[(1+\gamma-6h+2\sqrt{q})^{2}-(\gamma-1)^{2}]-\pi.\label{eq:BerryPhase2}\end{equation}

Now, let's analyze the connection between A-A phase and Berry phase
further. Above all, the condition of the adiabatic theorem reads\begin{equation}
\left|\frac{\langle m|\frac{d}{dt}H(t)|n\rangle}{E_{n}-E_{m}}\right|\ll1.\label{eq:AdiabaticThoerom}\end{equation}
 Substituting Eq. \eqref{eq:InstantaneousEigenEquation} into above
Eq. \eqref{eq:AdiabaticThoerom}, one can simplify the above condition
to\[
\omega\left|\langle m|\sigma_{z}|n\rangle\right|\ll1.\]
 By a simple calculation, one can know $\left|\langle m|\sigma_{z}|n\rangle\right|\sim1$,
so the condition becomes \begin{equation}
\omega\ll1.\label{eq:ConditionOfAdiabaticTheorom}\end{equation}
 Substituting the above Eq. \eqref{eq:ConditionOfAdiabaticTheorom}
into Eq. \eqref{eq:A-APhase1} and Eq. \eqref{eq:A-APhase2}, A-A
phase can be reduced to Berry phase which are Eq. \eqref{eq:BerryPhase1}
and Eq. \eqref{eq:BerryPhase2}. Moreover, for the degenerate A-A
phase of the second model, the readers who would like to verify it
will get the similar conclusion.

To sum up, we have analyzed both the non-degenerate and degenerate
A-A phase of LMG model. And the A-A phase have been reduced to corresponding
Berry phase according to the condition of quantum adiabatic theorem.
Furthermore, in order to calculate degenerate geometric phases, the
Floquet theorem and decomposition of operator have been generalized.
The general formula has been achieved.

This work was supported in part by NSF of China (Grants No.10605013
and No.10975075), and the Fundamental Research Funds for the Central
Universities.

\end{document}